\documentclass[preprint,12pt,times]{elsarticle}

\usepackage{amssymb}
\usepackage[numbers]{natbib}
\usepackage{amsmath}

\usepackage{float}
\usepackage{enumerate}
\usepackage{soul}
\usepackage{xcolor}
\usepackage{graphicx}
\usepackage{amssymb}
\usepackage{hyperref}
\usepackage{listings}
\usepackage{orcidlink}
\usepackage{tikz}
\usepackage{ulem}
\usepackage{aas_macros}
\usepackage{lineno}

\journal{Science China}

\begin{document}

\soulregister{\cite}{7}
\soulregister{\ref}{7}
\soulregister{\citep}{7}
\soulregister{\citet}{7}

\begin{frontmatter}



\title{The Spontaneous Genesis of Solar Prominence Structures Driven by Supergranulation in Three-Dimensional Simulations}


\author[1]{Huanxin Chen} 
\author[1,2,3]{*Chun Xia} 
\ead{chun.xia@ynu.edu.cn}
\author[1,2]{*Hechao Chen} 
\ead{hechao.chen@ynu.edu.cn}

\affiliation[1]{organization={School of Physics and Astronomy, Yunnan University},
            addressline={}, 
            city={Kunming},
            postcode={650500}, 
            state={Yunan},
            country={China}}
            
\affiliation[2]{organization={Yunnan Key Laboratory of Solar Physics and Space Science},
            addressline={}, 
            city={Kunming},
            postcode={650216}, 
            state={Yunan},
            country={China}}

\affiliation[3]{organization={National Astronomical Observatories, Chinese Academy of Sciences},
            addressline={}, 
            city={Beijing},
            postcode={100101}, 
            state={Beijing},
            country={China}}

\begin{abstract}
Solar prominences usually have a horizontally elongated body with many feet extending to the solar surface, resembling a multi-arch bridge with many bridge piers. The basic mechanism by which solar prominences acquire these common structures during their evolution, however, remains an unresolved question. For the first time, our three-dimensional magneto-frictional simulation, driven by supergranular motions, self-consistently replicates the commonly observed multi-arch bridge morphology and its characteristic structures of solar quiescent prominences in a magnetic flux rope. In comparison with traditional views, our simulations demonstrate that the spine, feet, and voids (bubbles) are inherent prominence structures spontaneously forming as the flux rope evolves to a mature state. The voids mainly consist of legs of sheared magnetic loops caused by unbalanced supergranular flows, and prominence feet settle at the bottom of helical field lines piled up from the photosphere to the spine. Similarities between the simulated prominences and observed real prominences by the Chinese H$\alpha$ Solar Explorer, the New Vacuum Solar Telescope, and NASA's Solar Dynamics Observatory suggest the high validity of our model. This work corroborates the pivotal role of photospheric supergranulation as a helicity injection source in the formation and shaping of quiescent prominence structures within the solar atmosphere, thereby paving a new avenue for future investigations into their fine dynamics and stability.
\end{abstract}


\end{frontmatter}


\section{Introduction}
\label{sec:intro}
Solar prominences are remarkable magnetized plasma structures commonly observed in the lower solar corona above polarity inversion lines (PIL)of photospheric magnetograms. Large quiescent prominences can persist for weeks or even months far away from active regions, and they are significant progenitors of coronal mass ejections. The magnetic topology of quiescent prominences is generally thought to be a helical magnetic flux rope (MFR) \citep{baksteslicka2013,Xia2014b,Xia2014a,Xia2016,Ouyang2017}, in which cool dense prominence plasma is supported against its gravity and thermally shielded from the much hotter corona. In observations, a typical quiescent prominence has a horizontally elongated body, named as a spine, with many feet extending to the solar surface, resembling a multi-arch bridge with many bridge piers. Bubble-like voids are frequently observed between prominence feet under the spine. As basic components, prominence feet are crucial for understanding the formation and stability of prominences, but their magnetic nature and genesis remain elusive.

When observed against the solar disk, prominences are also known as filaments, comprising a slender spine and a few bifurcated lateral extensions known as barbs \citep{Parenti2014}. The terms "filament" and "prominence" are often used interchangeably. Filament barbs and prominence feet are often regarded as two facets of the same structure \citep{Gibson2018,Chen2020}. In the early morphology-based models \citep{Martin1994b,Lin2008}, filament barbs were considered inclined magnetic field lines deviating from the spine, with their endpoints rooted at parasitic polarities, which are small patches of magnetic flux with polarity opposite to the main flux. On the other hand, force-free field models showed that barbs can be piles of magnetic dips, which are concave upward magnetic field line segments, extending from the spine, caused by the intrusion of small parasitic polarities near an MFR \citep{Aulanier1998,Mackay2009,Gunar2015a}. Moreover, prominence feet were also regarded as prominence tornadoes \citep{Pettit1932,Pettit1943}. These tornado-like feet often exhibit apparent rotational motions \citep{Suarez2012,Levens2015,Yang2018}. Some researchers \citep{Su2012,Wedemeyer2013,Su2014} intuitively interpreted the feet as rotating vertical magnetic structures. However, prominence feet are dominated by horizontal magnetic fields \citep{Schmieder2014b,Suarez2014,Levens2016a} and Doppler velocity measurements of long-term spectral observations excluded the possibility of rotational motions \citep{Martinez2016,Schmieder2017,Warren2018}. Hence, the apparent foot rotations are just illusions caused by horizontal oscillations within magnetic dips \citep{Panasenco2014,Gunar2023}.

Voids or bubbles underneath prominences appear as dark cavities in chromospheric lines but bright areas in coronal extreme ultraviolet (EUV) lines \citep{Stellmacher1973,Labrosse2011,Heinzel2008,Berlicki2011,Berger2017}. High-resolution observations revealed that bubbles could be real voids of cold prominence plasma \citep{Dudik2012}, in which a stronger EUV emission came from the foreground or background corona \citep{Berger2011,Shen2015,Wang2022} and the magnetic field strength inside small bubbles below prominences was higher than the surrounding prominence in spectropolarimetric observations \citep{Levens2016b}. Previous studies generally believe that prominence bubbles are caused by small magnetic bipoles beneath the prominences \citep{Mackay2009,Dudik2012,Gunar2014,Shen2015,Gunar2018}. By inserting a magnetic bipole under a prominence MFR, an arch-like magnetic field pushes away the MFR and prominence material, forming a relatively stable bubble. This scenario has been widely used to explain the appearance of prominence bubbles in many observations \citep{Dudik2012,Gunar2014,Shen2015,Berger2017,Guo2021}. There are two types of prominence bubbles in observations: transient bubbles in hedgerow quiescent prominences composed of vertical fine threads, and long-lived bubbles between feet in quiescent prominences composed of horizontal fine threads. We focus on the latter in this paper.

Although all the above studies suggest that the existence of small magnetic bipoles near prominences is a necessary precondition for the formation of prominence feet and voids, the rationale behind this hypothesis has yet to be confirmed in quiescent regions. In the filament channels of large-scale prominences in quiescent regions far from active regions, parasitic polarities are often small, very weak, and disorganized, giving rise to small-scale loops with disordered connectivity that exert little influence on the overall large-scale coronal magnetic field. Therefore, this parasitic-polarity-based assumption may not be applicable in all scenarios. Whether disorganized small parasitic polarities can produce a semi-regular distribution of feet along PILs and create voids at sufficient heights remains uncertain and likely depends on additional factors. To date, comprehending the three-dimensional magnetic nature of prominence feet and voids solely from observations is still extremely difficult. This is primarily due to the inherent difficulties in accurately measuring the coronal magnetic fields, despite recent progress in this area \citep[e.g.,][]{Yang2020,Yang2024}. 

Horizontal flows of supergranules in and out of a filament channel were measured by the coherent structure tracking \citep{Rieutord2007} applied to photospheric continuum images and they are similar \citep{Schmieder2014a}. A statistic study on the photospheric horizontal flows below filaments measured by local correlation tracking on Dopplergrams found only normal supergranular flows and no systematic converging or diverging flows with respect to the filament axis \citep{Ambroz2018}. The projected ends of filament barbs are located at the convergence maxima of the horizontal flows \citep{Schmieder2014a,Ambroz2018}. Stereoscopic observations based on 171 \AA images and magnetograms found that over 90\% of filament footpoints are located near supergranular boundaries \citep{Zhou2021}. These observational results imply an existing relationship between supergranules and filaments. A recent numerical model of quiescent prominence magnetic fields \citep{LiuandXia2022} attributes the prominence formation to supergranulations, which are convection cells of horizontally diverging flows sinking as anticyclonic flows at the cell boundaries due to the Coriolis force \citep{Hathaway1982,Duvall2000,Egorov2004,Hathaway2012,Langfellner2015}. Using this model, we successfully reproduced the formation of filament magnetic fields near solar poles \citep{Chen2024} by supergranular magnetic helicity injection and helicity condensation \citep{Antiochos2013}. Here, we further advance this model and find realistic prominence feet and voids in a mature stage. After confronting these numerical models with high-quality observations, we believe that supergranulations play a key role in shaping the structure of prominences.

\section{Method and Data} \label{sec:method}
\subsection{Simulation Methods}
We use two spherical wedge domains for the mid-latitude models ($1R_\odot<r<1.5R_\odot$, $17^\circ<\theta<53^\circ$, $0^\circ<\phi<60^\circ$) \citep{LiuandXia2022} and for the high-latitude models ($1R_\odot<r<1.5R_\odot$, $39.^\circ6<\theta<90^\circ$, $0^\circ<\phi<90^\circ$) \citep{Chen2024}, respectively. Each domain is discretized into five-layer adaptive mesh refinement (AMR) spherical grids with a logarithmic stretch in the radial direction \citep{Xia2018} and an effective resolution of $1024\times512\times512$ cells. Starting from bipolar photospheric magnetograms with quasi-elliptic smooth flux distributions using Gaussian functions, we extrapolate the initial potential magnetic fields using PDFI\_SS software \citep{Fisher2020}. Using the MPI-AMRVAC \citep{Keppens2023}, the magnetofrictional simulations are conducted by numerically solving the ideal magnetic induction equation $\frac{\partial \mathbf{B}}{\partial t}=\nabla\times(\mathbf{v}\times\mathbf{B})$ in which $\boldsymbol{v}=\boldsymbol{J}\times\boldsymbol{B}/(\nu_0 B^2)$ is the magnetofrictional velocity with the viscous coefficient $\nu_0 = 10^{-15}\ \text{s}\ \text{cm}^{-2}$ and $\boldsymbol{J}=\nabla\times\mathbf{B}/\mu_0$. The velocity smoothly decays to zero toward the photosphere \citep{Cheung2012}, with an upper limit of 30 km s$^{-1}$ \citep{Pomoell2019}. As magnetic reconnection occurs due to numerical resistivity, the improved resolution in our simulation—compared to \citet{LiuandXia2022,Chen2024}—reveals more magnetic features. We use the constrained transport scheme \citep{Gardiner2005} on a staggered AMR mesh \citep{Olivares2019} to ensure zero divergence of the magnetic field. Boundary conditions are set as periodic on the longitudinal boundaries, open on the outer radial boundary, and closed on the latitudinal boundaries. On the photospheric boundary, we impose a zero radial velocity and a supergranular horizontal velocity field, which is described as follows.

A Voronoi tessellation of the solar surface is used to resemble the topological and statistical characteristics of supergranular cells \citep{Schrijver1997}. The center points of supergranular cells are evenly distributed using Poisson disk sampling with a minimum interval of 20 Mm. The area of supergranular cells is proportional to a time-dependent weight function $\omega=3|\sin(\pi t/\tau+\xi)|+0.7$, in which $\tau$ represents the lifetime of a supergranular cell with a Gaussian random distribution around 1.6 days \citep{Hirzberger2008}, and $\xi$ is the random initial phase. A supergranule is removed from the tessellation if its weight falls below 1, and it is regenerated if its weight grows above 1. The dimensionless horizontal diverging velocity within each supergranular cell is given by $v_r(r)=2r^2/r_0\exp(-4r^2/r_0^2)$, where $r$ is the spherical distance to the center of the cell, and $r_0$ is the radius of a circle whose area is equal to that of the supergranular cell \citep{Gudiksen2005}. The dimensionless horizontal rotating velocity, as a result of the Coriolis force acting on the diverging horizontal flows, is set as $v_t(r)={2r/r_0}v_r(r)$ with additional latitude dependence of the Coriolis force only for the high-latitude models. We use a scale factor to set the maximum supergranular velocity to be 500 m s$^{-1}$. We also add the velocities of differential rotation and meridional flow \citep{Mackay2014} to the horizontal velocity. Then, we multiply the total photospheric driving velocity by 5 to speed up the long-term evolution; even after this enhancement, the velocity remains within the observed range of 0.5–2.5 km/s \citep{Rincon2018}.

\subsection{Observational Data}
The observational data are from the New Vacuum Solar Telescope (NVST) \citep{Liu2014,Yan2020} in China, the Chinese H$\alpha$ Solar Explorer (CHASE) \citep{Li2022}, and the Solar Dynamics Observatory (SDO) \citep{Pesnell2012}. The H$\alpha$ Imaging Spectrograph (HIS) onboard CHASE conducts full-disk spectroscopic observations in the H$\alpha$ (6559.7–6565.9~\AA) wavebands, using raster scanning mode. HIS provides a spectral resolution of 0.024~\AA~per pixel, a spatial resolution of $0.^{\prime \prime} 52$ per pixel, and a temporal resolution of 60 seconds. For our analysis, we utilized the H$\alpha$ line center (6562.8~\AA). The high-resolution H$\alpha$ center images of the prominence recorded by the NVST were used in this study. These H$\alpha$ images have a field of view of $150^{\prime \prime} \times 150^{\prime \prime}$, a cadence of 12 seconds, and a CCD plate scale of $0.^{\prime \prime}165$ per pixel. The line-of-sight magnetogram data we used came from the Helioseismic and Magnetic Imager (HMI) on board the SDO, with a temporal cadence of 720 seconds (hmi.M\_720s). 

\section{Results}

\subsection{Formation of Magnetic Flux Rope}

Magnetic dips, which form gravitational potential wells, are stable locations for dense prominence plasma to stay or oscillate around. So we treat magnetic dips as approximate representatives of prominence plasma structures \citep{DeVore2000,Chen2022}. We define magnetic dip regions as cells with positive radial curvature of the magnetic field and $<10 \%$ proportion of the radial component of the magnetic field. 
Figure~\ref{fig:fig1} illustrates the formation process of feet and voids in the high-latitude prominence model. At 150 hr, due to the motion of supergranules, the magnetic field arcades along the PIL exhibit a quasi-periodic potential-sheared-potential pattern (Figures~\ref{fig:fig1} (a1), (b1)). At 450 hr (Figures~\ref{fig:fig1} (a2), (b2)), multiple short MFRs and corresponding isolated magnetic dip clumps appear along the PIL, with the axes of the MFRs already connected. By 800 hr (Figures~\ref{fig:fig1} (a3), (b3)), the small MFRs have grown and merged into one long MFR, and most of the dip clumps have merged into a slab structure. At this time, the magnetic dips of the MFR are approximately 15 Mm in height. By 1300 hr (Figures~\ref{fig:fig1} (a4), (b4)), the dip regions of the MFR have grown to about 28 Mm in height, with small voids gradually emerging from the bottom of the dip region, where the underlying PIL parts are curved away from the filament spine. At 1650 hr (Figures~\ref{fig:fig1} (a5), (b5)), several large voids appear, and the height of dip regions reaches approximately 50 Mm. An animation of Figure~\ref{fig:fig1} is provided. There are two types of void formation processes marked by the red box and the blue box in Figures~\ref{fig:fig1} (a5) and (b5). In the red box, unlike foot regions where continuous magnetic flux cancellation connects footpoints of sheared arcades and creates magnetic dips joining the foot at the bottom, the uneven supergranular motions push the local PIL to protrude to one side of the filament spine (as seen in Figure~\ref{fig:fig5} (c1)), stop flux cancellation, and stretch the legs of sheared arcades to the side to form a void without magnetic dips. In the blue box, the right dip clump connects with the left dip region at a large coronal height, forming a void underneath via coronal magnetic reconnection between two bundles of helical field lines of the MFR. 

We have done numerous experiments with different supergranule distributions at different latitudes. The exact locations of feet and bubbles/voids vary, but the quasi-periodic foot-bubble-foot distribution pattern persists. All simulations can reach a steady state after the bottom boundary driving velocity is turned off. In future models including plasma, the gravity of prominence material can be balanced by the upward magnetic tension force in the dip regions of the MFR, allowing the system to naturally reach a dynamic equilibrium.

\subsection{Mature Magnetic Flux Rope Model}

We illustrate an overview of the mature prominence MFR in Figure~\ref{fig:fig2}, which exhibits an inverse S-shaped sigmoid and possesses negative magnetic helicity, in the high-latitude simulation. Overall, the simulation successfully reproduces the commonly observed multi-arch bridge morphology of solar quiescent prominences, consisting of horizontal spine, vertical feet, and voids (Figures~\ref{fig:fig2} (a), (d), and (e)). The colored magnetic dips delineate regions capable of hosting prominence plasma. The mature MFR features a continuous region of dips (in red) over 30 Mm in height, which represents a spine, and several pillar-like dip regions (in yellow-green-blue), which represent prominence feet (Figures~\ref{fig:fig2} (a) and (d)). Under the spine and between the feet, bubbles as voids of magnetic dips become apparent in the mature stage. Magnetic field lines contain magnetic dips in the spine and feet are mainly helical field lines, and the voids contain inclined legs of arch field lines that are devoid of dips and situated above the MFR (as shown in Figure~\ref{fig:fig2} (a); see also Figure~\ref{fig:fig5}). In panel (a), localized weak fields at the centers of some supergranules cause the overlying field lines to bend downward, producing very low-lying magnetic dips (shown in dark blue). This effect is limited to field lines at chromospheric heights and does not modify the magnetic structure of the prominence. The higher magnetic dips are closer to the axis of the MFR and exhibit stronger magnetic field strength than the lower dips, which is consistent with the magnetic field measurements of limb prominences \citep{Schmieder2014b,Suarez2014,Levens2016a}. In Figure~\ref{fig:fig2} (c), the bottom photospheric magnetogram well mimics a real photospheric magnetic network, which is primarily characterized by a network of sporadic magnetic flux concentrations at supergranular boundaries. Note that parasitic polarities, small magnetic fluxes with polarity opposite to the surrounding main flux, are not included in our models at all. 

Magnetic field line segments in magnetic dip regions in Figures~\ref{fig:fig2} (d)-(f) can host dense cool plasma and approximately represent fine filament threads. Low-lying magnetic field line segments, starting from the strong magnetic flux regions, can dynamically host dense cool plasma and represent chromospheric fibrils. In the side view (Figure~\ref{fig:fig2} (d)) of the prominence above the limb, the feet (shown in yellow-green-blue) extending from the chromospheric height up to the spine, with the apparent width increasing with height, closely resemble the pillar-like feet of real prominences, and this configuration persists under all views, as also reported in previous studies from the side view \citep{Aulanier1998,Aulanier1998b}. In the top view of the filament against the solar disk (Figure~\ref{fig:fig2} (e)), the same feet branch out from the spine, forming right-bearing barbs, highlighted in the yellow boxes, consistent with the negative helicity of typical dextral filaments \citep{Martin1998}. The chromospheric fibrils, radiating from strong magnetic flux concentrations near the filament, stream align with the PIL towards opposite directions on the two sides of the PIL, consistent with the observed H$\alpha$ fibril patterns \citep{Martin1998}. When the magnetic field lines of the chromospheric fibrils extend into the corona, they may form the coronal cell \citep{Sheeley2013}, namely Fe $\textrm{XII}$ stalks \citep{Wang2013}, with a counterclockwise whorl and a dextral chirality, in which case the coronal cells and the chromospheric fibrils on the right (left) side of the PIL point away from (toward to) an observer looking along the PIL. When the modeled prominence comes to the west solar limb, as shown in Figure~\ref{fig:fig2} (f), the prominence voids are not distinctly visible. Instead, the prominence appears as a cohesive pillar structure with descending flanks \citep{Berger2011}, which are attributed to the overlapping of many dipped field line segments. This simulated prominence highly resembles most morphological features of a real solar prominence from different views.

\subsection{Comparison to Prominence Observations}

To better compare our simulations with actual prominence observations, we conducted a series of observations on a quiescent prominence for one week since November 21, 2022, utilizing the NVST in China and the CHASE satellite. Figures~\ref{fig:fig3} (a1)-(a3) and (b1)-(b3) depict our sequential tracking observation of the prominence as it transited from the eastern solar limb to the solar disk center. At 05:41 UT on November 21, the observed prominence exhibited three feet, two voids between the feet, and a dim spine connecting the feet (Figure~\ref{fig:fig3} (b1)). About eight hours later, as the prominence partially rotated towards the solar disk, its feet gradually transformed into corresponding barbs (Figure~\ref{fig:fig3} (a1)). By November 23, the prominence became an on-disk filament characterized by three barbs and two voids (Figures~\ref{fig:fig3} (a2) and (b2)). On November 26, as the filament transited with solar rotation towards the center of the solar disk, its spine became increasingly discernible, but its feet became shorter and less discernible, likely due to the effects of projection and obscuration (Figures~\ref{fig:fig3} (a3) and (b3)). The enhanced visibility of the spine may be attributed to mass redistribution, whereas the three latitude lines show that the latitudinal positions of the stable feet remain largely fixed.

Given the relatively slow evolution of the magnetic skeleton of quiescent filaments, we can utilize a simulated filament rotated into different angles to compare with the observed prominence tracked for several days. In Figures~\ref{fig:fig3} (c1)-(c3), we present zoom-in snapshots of the magnetic dip regions of a simulated mid-latitude filament at 1495 hr, captured from similar viewing angles as the observations. Above the solar limb, this simulated magnetic-dip prominence exhibits classic structures, including a spine (in red), feet (in yellow-green-blue), and two stable voids between feet. The magnetic dips can more stably support cool plasma with larger optical depth, such as the pillar-like feet observed in H$\alpha$ and 171~\AA. From the solar limb to the disk center, one can see that the morphology evolution of the simulated prominence demonstrates a striking similarity with the real prominence. 

\subsection{Fragmented Quiescent Filaments}

In on-disk observations, the growing of quiescent filaments commonly begins with several fragmented filament clumps located at the boundaries of supergranular cells \citep{Pevtsov2005}. These isolated clumps may subsequently grow and connect with each other, and eventually become a coherent filament. In March 2018, the NVST captured such a formation process of a quiescent filament, as depicted in Figures~\ref{fig:fig4} (b1)-(b3). Initially, four isolated filament clumps appeared at 06:26 UT. About 2.5 hours later, these filament clumps gradually extended and connected to form an elongated quiescent filament. To date, the underlying mechanism behind this common phenomenon remains elusive.

In the early phase of all our simulations, a very similar formation process of quiescent filaments, from fragmented to coherent, is approximately reproduced. As shown in Figures~\ref{fig:fig4} (a1)-(a3), four isolated magnetic dip clumps appear early at 300 hr. These magnetic dip regions expand and grow along the PIL, with the two southern ones merged at 600 hr. By 900 hr, the segmented dip regions coalesce into a coherent filament. In fact, the isolated filament clumps in the simulations, which are the first-appearing magnetic dips in the lower parts of short small MFRs, are the seeds of filament feet, which may develop into mature feet of coherent filaments. Filament plasma may form and first accumulate in these segmented dip regions, naturally giving rise to isolated filament clumps. Despite the difference in time scale, the simulation results successfully replicate the observed NVST quiescent filament formation process — from fragmented clumps to a unified filament — in the first approximation.

Combining our simulation results \citep{Chen2024}, we propose a filament chirality rule based on the PIL shape under the fragmented filaments: in the northern (southern) hemisphere, these isolated filament clumps preferentially form above the S-shape (z-shape) PIL parts of dextral (sinistral) filament channels (Figures~\ref{fig:fig4} (c1), (c2)). In Figures~\ref{fig:fig4} (d1) (also see \citet{Filippov2017}) and (d2), we include two representative fragmented H$\alpha$ quiescent filaments to validate the proposed empirical rule. By plotting the smoothed large-scale PIL of each filament, it becomes clear that fragmented filaments with dextral/sinistral patterns indeed emerge along the curved S-shape/Z-shape PIL parts, as our rule predicted. Therefore, we suggest that this so-called ``S-Z PIL rule" has the potential to predict the sites of the first appearance of filament clumps by examining the morphologies of their corresponding PILs in real observations.

\subsection{Magnetic Nature of Feet and Bubbles}

Figure~\ref{fig:fig5} a1 - c2 depicts the magnetic field lines of different structures of the high-latitude prominence model, including spine, feet, and voids, in a mature MFR (at 1705 hr) from top views and side views. Magnetic field lines traversing the spine, feet, and voids are delineated in red, yellow, and orange, respectively. The spine consists of the magnetic dips of helical field lines with inverse S shapes constituting the inner layers of the MFR. The helical field lines of the middle part of the spine are more symmetric than the ones of the two end parts. Regarding prominence feet, they indeed consist of a collection of magnetic dips, but they are not the result of the intrusion of small-scale parasitic polarities that are commonly believed in previous studies \citep{Aulanier1998,Gunar2016,Shen2015}. The yellow twisted magnetic field lines of the feet are similar to those of the spine, albeit located at the outer layers of the MFR (Figures~\ref{fig:fig5} (b1), (b2)). Panel (b1) shows the supergranular cells in blue lines. The colored feet are spatially correlated, in projection, with the supergranular boundaries, consistent with the observations. The voids under the prominence are occupied by inclined legs of sheared magnetic arcades (in orange) overlying the MFR, and minor locally-arched helical field lines. These magnetic arcades associated with the voids interweave beneath the MFR and are intricately entangled with it. Unlike previous studies \citep{Dudik2012}, the voids are not caused by small bipolar magnetic loops under the MFR. One end of the orange magnetic field line is directly anchored beneath the MFR, while the other end is connected to a distant location, or both ends lie near the MFR. By displaying more magnetic field lines in the void and foot region, it is evident that most void lines belong to the overlying magnetic arcade, whose legs intrude beneath the MFR (Figures~\ref{fig:fig5} (d1)-(d2)). After inspecting the evolution of the model, we find that the strong deformation of the PIL under the influence of supergranulations accompanies one-sided protrusions of inclined legs of overlying magnetic arcades, causing the formation of voids. Most foot field lines belong to the MFR; they correspond to the upward-concave field lines (Figures~\ref{fig:fig5} (d3)-(d4)) \citep{Chen2025}. Therefore, the MFR itself naturally develops a magnetic field topology characterized by a quasi-periodic ``foot-bubble-foot" pattern as its evolution proceeds. In our simulations, this strongly suggests that the feet and bubbles of prominences are intrinsic evolutionary results when the prominences evolves to a mature state, rather than a deformation/distortion of the pre-existing prominence magnetic-field structures in response to the intrusion of external parasitic polarities \citep{Dudik2012} or the flux imbalance of ambient magnetic fields \citep{Van2004}, as proposed in previous prominence models.

Furthermore, we measured the mean magnetic field strength of the spine, feet, and voids along the mature prominence MFR. Our rough measurements suggest that: (1) the feet region is dominated by a horizontal magnetic field rather than a vertical \citep{Levens2016b}; and (2) a similar mean magnetic field strength ($\sim$ 8.124 Gauss) exists inside the void region than in the surrounding prominence structure ($\sim$ 8.274 Gauss).

\section{Conclusion and Discussion} \label{sec:DS}

Through the improvements and deep analysis of our supergranule-driven prominence magnetic field models \citep{LiuandXia2022,Chen2024}, we present the first self-consistent three-dimensional simulations of the formation and evolution to a mature state of quiescent solar prominences. This new prominence model only imposes supergranular flows on the photosphere self-consistently to generate the magnetic field configuration of a common long prominence. This model successfully replicates the commonly observed multi-arch bridge morphology and its characteristic structures of solar quiescent prominences for the first time. It thus enables us to characterize the topological properties of characteristic prominence structures, including spine, feet, and voids, and to ascertain their origin during the simulated evolution of the prominence magnetic field. Meanwhile, the strong similarities between the simulated prominences and observations from CHASE, NVST, and SDO suggest the high validity of our model. 

Classic prominence models attributed the appearance of prominence feet to the distortion of a cylindrical magnetic flux rope by the addition of parasitic polarities \citep{Aulanier1998}. When a parasitic polarity is far from the flux rope axis, the feet and the spine (flux rope axis) are above two different PILs. As a parasitic polarity is placed closer to the axis, it merges into the main polarity on the other side with the same sign and the two PILs merges into one curved PIL \citep{Aulanier1998b}, which is the case very similar to our model in which the feet and the spine are above the same curved PIL but the parasitic polarity can no longer be distinguished. Both models propose that prominence feet are stacked magnetic dips above PILs with bald patches and they are not inclined field lines anchored in parasitic polarities.  The curved PIL and the lateral protrusion of the feet is due to local imbalance of converging flows from supergranules, which is different from the idea of imbalance of ambient magnetic fields \citep{Van2004}. The end points of prominence feet in our models are located near boundaries of supergranules, which is in accord with observations \citep{Schmieder2014a,Ambroz2018,Zhou2021}. Regarding voids, \citet{Gunar2014} argued that distinguishing bubbles by emission enhancement alone is difficult because foreground and background structures are hard to disentangle along the line of sight. Therefore, the magnetic configuration of bubbles is crucial for understanding their nature. All previous models manually inserted a small magnetic bipole under an MFR to create a bubble, which is threaded by a small magnetic arcade below the prominence \citep{Mackay2009,Gunar2014,Gunar2018}. In contrast, the voids in our models are threaded by the inclined legs of large sheared magnetic arcades overlying the prominence, which are mainly caused by unbalanced supergranular flows pushing magnetic flux across the MFR axis. Threaded by inclined field lines, the void regions are not able to stably contain dense prominence plasma for a long time. 

In comparison with traditional views, our simulations clearly demonstrate that parasitic polarities and small bipoles are not a prerequisite for the formation of prominence feet and voids. In our model, all commonly observed prominence structures are inherent complexities in an MFR spontaneously forming driven by supergranulation and magnetic helicity condensation. The alternating arrangement of feet and voids along the prominence at supergranular scales is influenced by supergranular flows and yet to be further confirmed by future observations. The magnetic fields of prominence feet consist of piled shallow dips of the helical magnetic field in the outer layer of the MFR as a result of magnetic flux cancellation driven by supergranular converging flows. Our results strongly corroborate that photospheric supergranulation as a basic helicity injection source plays a key role in the formation and shaping of quiescent prominence structures in the solar atmosphere.

Moreover, we explain the filament formation from fragments to a mature integral filament as the growth and integration of small MFRs. The feet, owing to their deeper and lower-lying magnetic dips, serve as favorable sites for plasma accumulation. We offer a ``S-Z PIL rule" to predict the sites where filament clumps first appear. The early clumps and later feet of dextral (sinistral) filaments are preferentially located above the S-shape (Z-shape) PIL parts.

\begin{figure*}[htbp]
\centering
\includegraphics[width=\textwidth]{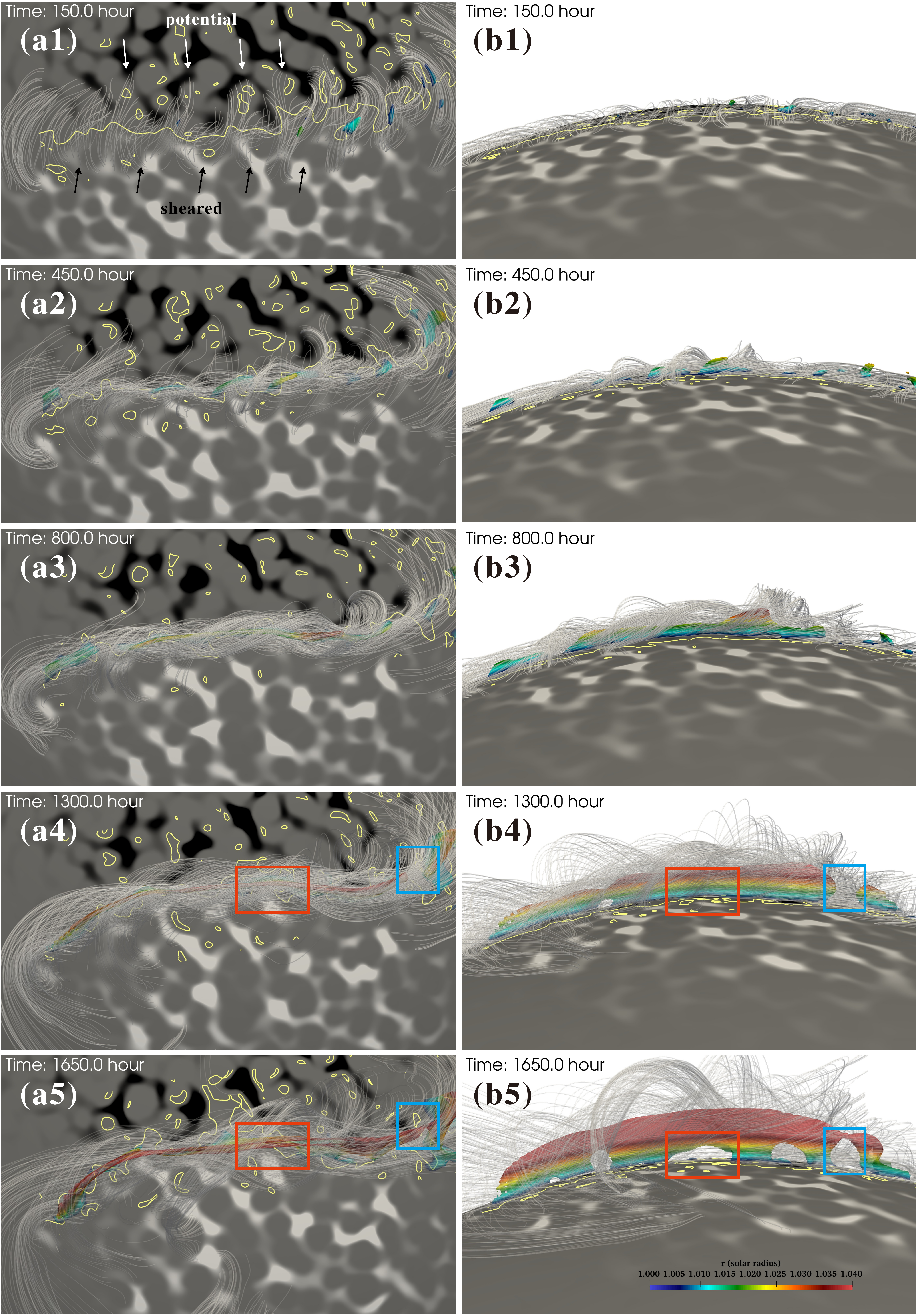}
\caption{The formation of prominence spine, feet and voids. The left and right columns depict a time series of top-view (panels (a1) - (a5)) and side-view (panels (b1) - (b5)) images, respectively. The colored lumps represent dip regions at 1.0055-1.1 R$_\odot$, with saturation beyond 1.04 R$_\odot$. The white lines are magnetic field lines passing through the uniformly sampled points on the PILs of spheres at 1.005, 1.01, 1.015, 1.02, and 1.04 R$_\odot$. The black and white arrows indicate the shear pattern of the magnetic field lines. The red and blue boxes mark two types of void formation. Photospheric magnetograms and PILs are shown at the bottom. An animation of this figure is available (see the Supplementary Materials).}
\label{fig:fig1}
\end{figure*}

\begin{figure*}[htbp]
\centering
\includegraphics[width=\textwidth]{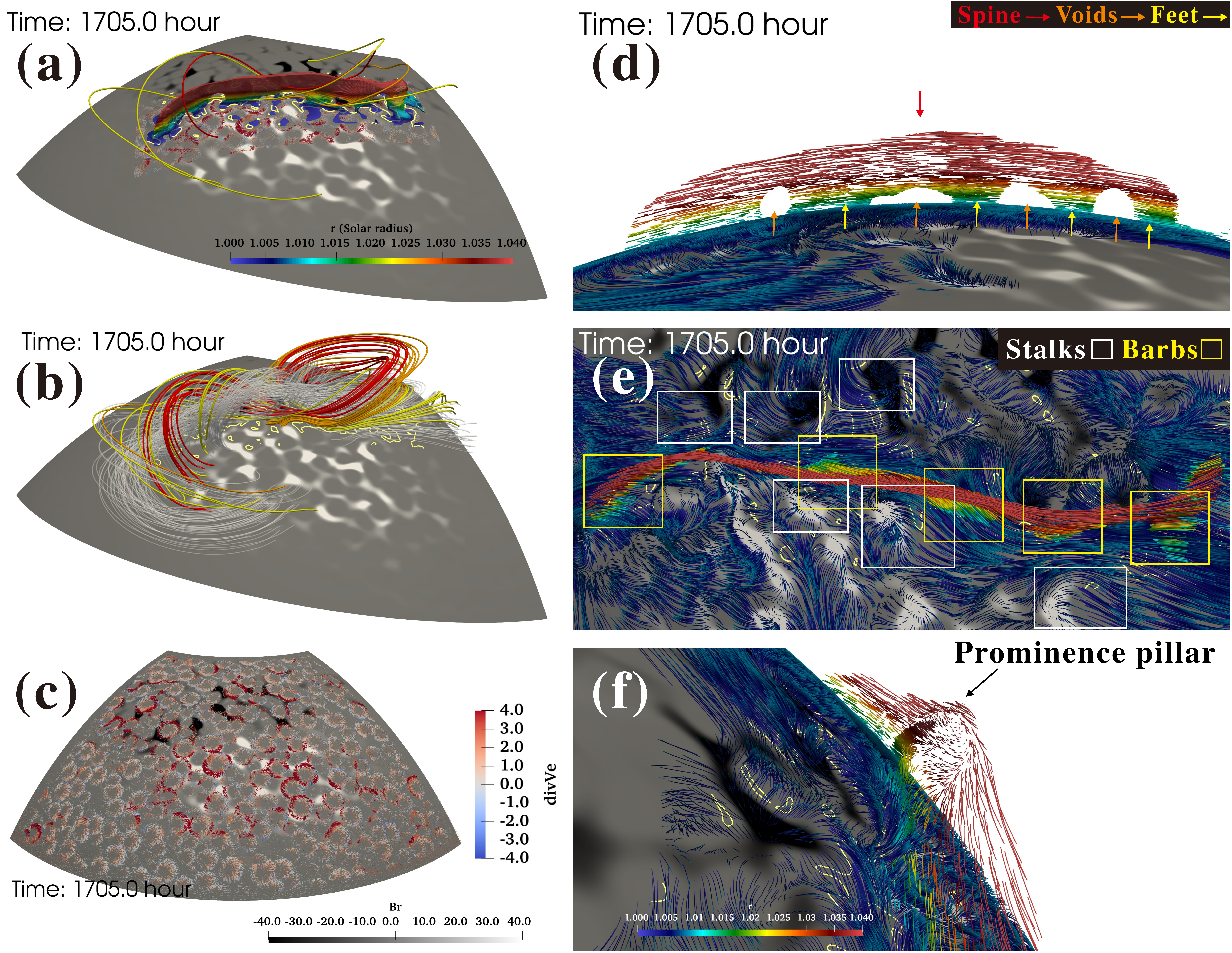}
\caption{The high-latitude prominence model from 1705-hour magnetofrictional simulation. (a), Isosurfaces representing magnetic dips are rendered using a rainbow color scheme, with colors mapped to the solar radius $r$. The red, yellow, and orange magnetic field lines pass through the spine, feet, and voids, respectively. The bottom photospheric boundary presents the white positive and black negative radial magnetic fluxes saturated at $\pm40$ G, the PILs in yellow, and horizontal supergranular flows in arrows near the MFR. (b), White lines depict the main body of the MFR, while red, yellow, and orange lines thread through the spine, feet, and voids, respectively. (c), Horizontal supergranular velocity vectors, colored by the divergence of velocity, with red indicating divergence and blue indicating convergence. In panels (d) - (f) with the same visualization in different viewing angles, the prominence is represented by magnetic field line segments within magnetic dip regions, field lines from the magnetic network are clipped beyond the transition region, resembling chromospheric fibrils, and all field lines are colored by $r$.} 
\label{fig:fig2}
\end{figure*}

\begin{figure*}[htbp]
\centering
\includegraphics[width=\textwidth]{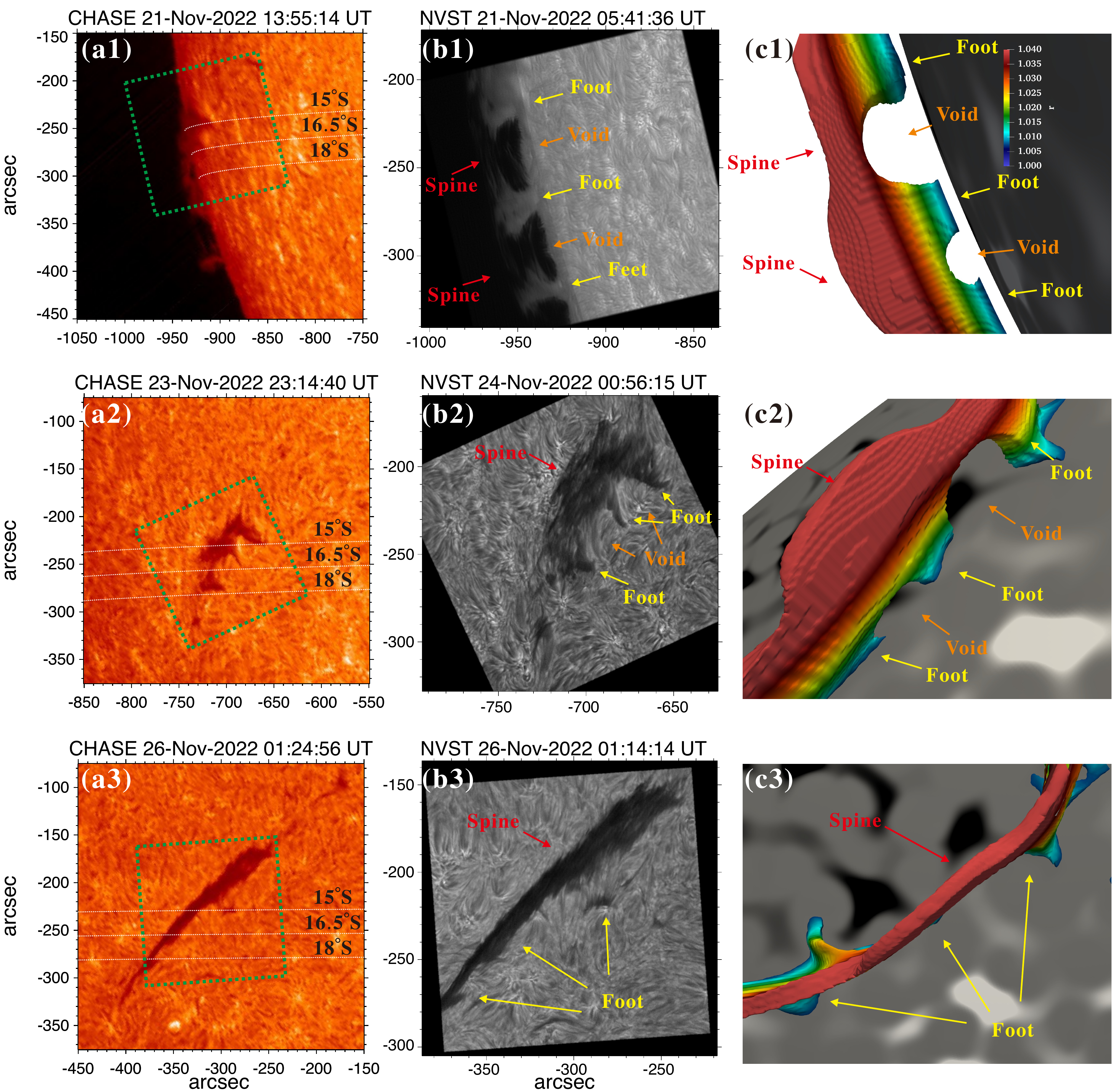}
\caption{Large-field-of-view H$\alpha$ images (a1) - (a3) from CHASE and high-resolution H$\alpha$ images (b1) - (b3) from NVST monitoring a quiescent prominence in a sequence of times and perspectives, with the dashed green boxes indicating the NVST field of view. The three dashed lines indicate the latitudes where feet are rooted, at $-15^\circ$, $-16.5^\circ$, and $-18^\circ$. The simulated prominence of the mid-latitude model at 1495 hr (c1) - (c3) is depicted from similar viewing angles as in (b1) - (b3). Red, yellow, and orange arrows indicate spines, feet, and voids in both the observed and the simulated filament. Isosurfaces depict magnetic dip regions of prominence from 1.0055 to 1.1 R$_\odot$, with height-indicating colors saturated at $1.04r_{\odot}$. The simulated photospheric magnetogram under the prominence is rendered in a grey scale.}
\label{fig:fig3}
\end{figure*}

\begin{figure*}[htbp]
\centering
\includegraphics[width=\textwidth]{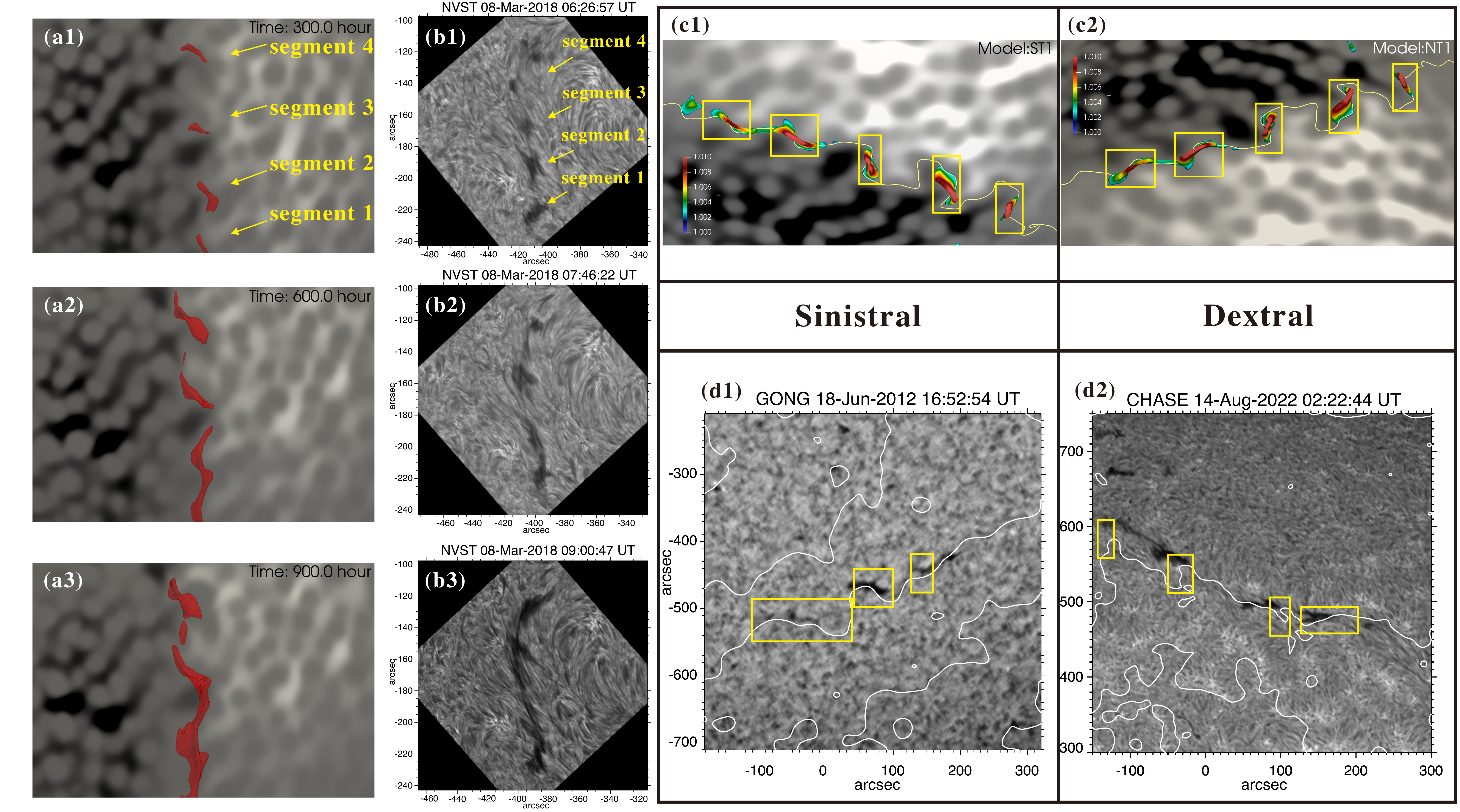}
\caption{The growth and connection of filament segments. Panels (a1) - (a3) present a similar filament formation process above magnetograms in the mid-latitude numerical model with red isosurfaces depicting magnetic dip regions. H$\alpha$ images (b1) - (b3) illustrate a filament formation process observed by NVST. Yellow arrows indicate the initial four filament segments in (a1) and (b1). High-latitude models (c1) and (c2) at 450 hr present fragmented magnetic dip regions above the yellow Z-shape and S-shape PIL parts on the photospheric magnetograms in the southern and northern hemispheres, respectively. H$\alpha$ image d1 shows a short fragmented filament with sinistral chirality above Z-shape PIL parts in the southern hemisphere, while (d2) presents a forming filament with dextral chirality above S-shape PIL parts in the northern hemisphere. The white lines in panels (d1) and (d2) are the large-scale PILs at low coronal heights of the potential field extrapolated from the HMI magnetograms.}
\label{fig:fig4}
\end{figure*}

\begin{figure*}[htbp]
\centering
\includegraphics[width=0.9\textwidth]{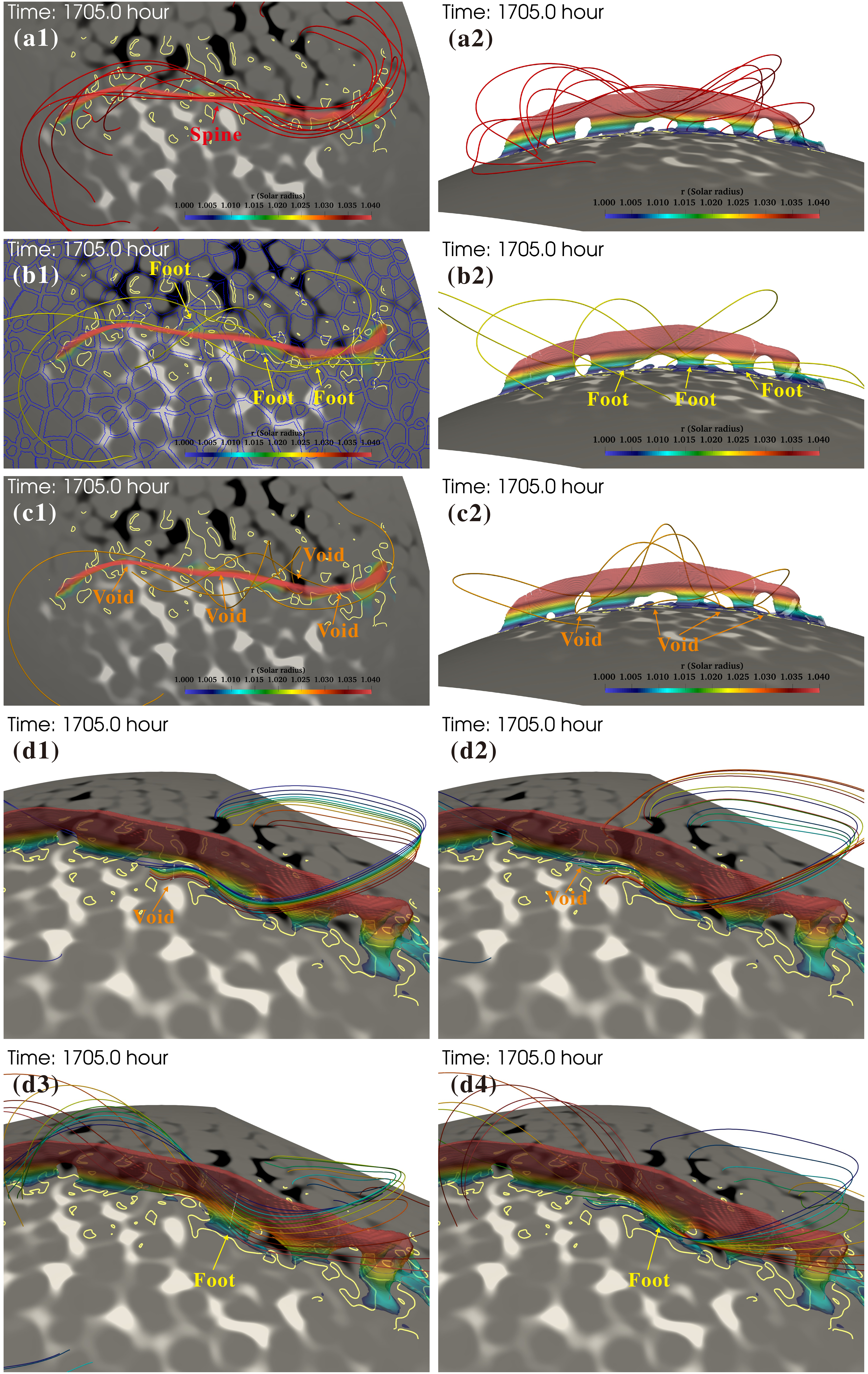}
\caption{Magnetic topology of spine, feet, and voids at 1705 hr in the high-latitude model. The volume rendering of the magnetic dip regions of the prominence has transparency to show the red, yellow, and orange magnetic field lines passing through the spine, feet, and voids, respectively, in the top views (left column, (a1) - (c1)) and the side views (right column, (a2) - (c2)). In panel (b1), the supergranular cells are described in blue. Panels (d1) - (d4) display particular magnetic field lines through the void and foot region. The white line indicates the reference line along which the magnetic field lines are sampled.}
\label{fig:fig5}
\end{figure*}

Real solar prominences are composed of fine structures known as filament threads. Observations show that these threads are short-lived, typically lasting only a few minutes to tens of minutes \citep{Lin2005}. In Figure~\ref{fig:fig2} (d)-(f), we represent these filament threads by segments of magnetic field lines located in the dip regions. A uniform sampling method was employed to plot the field lines, which resulted in a visual appearance of separated, thread-like structures. \citet{Zhou2020} demonstrated that the formation of a filament thread leads to the development of a low-pressure region, which induces the siphoning of surrounding plasma. This process causes the shrinkage of the magnetic flux tube containing the thread, while adjacent flux tubes tend to expand. This implies that once a thread is formed, its neighboring flux tubes become less favorable for further condensation. In Figure~\ref{fig:fig1}, it can be seen that the spatial distribution of magnetic field lines in the three-dimensional MFR driven by the supergranular velocity field is not entirely uniform. The magnetic field lines themselves are inherently non-uniformly distributed, and the modulation by the inhomogeneous velocity field may lead to a more complete separation of flux tubes than previously expected. In our simulations, the initial distribution of the ideal artificial magnetogram is relatively uniform, whereas real solar magnetograms are undoubtedly more complex. However, the randomly distributed supergranulations quickly transform the smooth magnetogram into a magnetic network, the initial smooth flux distribution is soon forgotten. Therefore, we will get a similar magnetic network and similar prominence MFR formation, if we start from a realistic magnetic work instead.

The updated supergranule-driven prominence magnetic model demonstrates significant potential in reproducing and topologically characterizing many key observational features of quiescent prominences, such as their stable spine, foot, and void morphology. It opens a new avenue for future investigations into prominence dynamics and stability in response to solar surface motions. Using this reliable model as a starting point, future studies will employ full magnetohydrodynamic (MHD) simulations to achieve better self-consistency and more realistic prominence models. A more realistic pressure scale height for both prominence plasma and the coronal atmosphere \citep{Low1982} will be incorporated. In the future, this basic model will enable us to reproduce the fine dynamics and instabilities inside solar prominences in response to solar surface motions or waves from distant explosions, revealing crucial insights into the coupling between plasma and the magnetic field within these structures. 


\section{Declaration of Interests}
The authors declare that they have no conflict of interest.

\section{acknowledgments}
This research was supported by the Strategic Priority Research Program of the Chinese Academy of Sciences (XDB0560000), 
the National Natural Science Foundation of China (12073022, 11803031), 
the Key Research Program of Frontier Sciences, CAS (ZDBS-LY-SLH013), 
Scientific Research and Innovation Project of Postgraduate Students in the Academic Degree of YunNan University (KC-23233895, KC-24248785), 
and the Yunnan Key Laboratory of Solar Physics and Space Science (202205AG070009). 
H.C.C. was supported by NSFC (12573061), Yunnan Key Laboratory of Solar Physics and Space Science (YNSPCC202210) 
and the Yunnan Provincial Basic Research Project (202401CF070165). 
The numerical simulations were conducted on the Yunnan University Astronomy Supercomputer. 
The NVST data were obtained from a topic observation led by C.X. and H.C.C. granted by Yunnan Observatories. 




\bibliographystyle{num}
\bibliography{reference}

\end{document}